\documentstyle[pra,aps]{revtex}
\begin{document}
\draft

\def\overlay#1#2{\setbox0=\hbox{#1}\setbox1=\hbox to \wd0{\hss #2\hss}#1%
\hskip -2\wd0\copy1} \twocolumn[
\hsize\textwidth\columnwidth\hsize\csname@twocolumnfalse\endcsname

\title{Reply to the ``Comment on `quantum backaction of optical observations
on Bose-Einstein condensates' ''}
\author{U.\ Leonhardt$^1$, T. Kiss$^2$, and P.\ Piwnicki$^3$}
\address{~$1$School of Physics and Astronomy, University of St Andrews,
North Haugh, St Andrews, Fife, KY16 9SS, Scotland}
\address{~$^2$Department of Nonlinear and Quantum Optics,
Institute for Solid State Physics and Optics, P.O. Box 49, H-1525
Budapest, Hungary}
\address{~$^3$Physics Department, Royal Institute of Technology (KTH),
Lindstedtsv\"agen 24, S-10044 Stockholm, Sweden} \maketitle
\date{today}

\vskip2pc] \narrowtext

How is a Bose-Einstein condensate perturbed by dispersive imaging
\cite{Andrews}? Suppose that dispersive imaging is absorptionless,
what is the quantum backaction?

In our paper \cite{LKP} we have only addressed the second
question. We found that two processes contribute to the
backaction, phase diffusion and condensate depletion. The
pioneering paper \cite{Andrews} referred only to phase diffusion
as the quantum backaction and made no estimation of the diffusion
rate. According to our calculation \cite{LKP}, depletion turned
out to dominate the quantum backaction. Since the depletion rate
agreed with the experimental facts \cite{Andrews} within the
available accuracy, we conjectured that the quantum backaction
might indeed explain the observed perturbation. However, in our
analysis we have entirely ignored the residual absorption of
dispersive imaging, in order to determine the backaction {\it per
se}. The Comment \cite{Comment} stresses the fact that absorption
is still stronger than the quantum backaction. This is correct,
as can be seen from the following calculation.

The depletion rate due to backaction is, according to Eq.\ (62)
of Ref.\ \cite{LKP},
\begin{equation}
\gamma_L= \frac{\pi^2}{4} \frac{\chi_0^2}{\hbar c} I \lambda^{-3}
\,,
\end{equation}
where $\lambda$ denotes the wave length of light, $I$ is the
intensity and $\chi_0$ is the susceptibility per atom-wave
density. We assume the individual atoms as two-level systems with
detuning $\Delta$ and Rabi frequency
\begin{equation}
\omega_R = \frac{d}{\hbar}\, E^{(+)} \,,
\end{equation}
where $d$ is the dipole moment and $E^{(+)}$ is the positive
frequency part of the electric field strength. Our goal is to
compare $\gamma_L$ with the spontaneous emission rate \cite{SZ}
\begin{equation}
\Gamma=\frac{1}{4\pi\epsilon_0}\,\frac{4d^2}{3\hbar}\,
\left(\frac{2\pi}{\lambda}\right)^3=
\frac{1}{4\pi\epsilon_0}\,\frac{4d^2\hbar\,|\omega_R|^2}
{3\,|E^{(+)}|^2}\, \left(\frac{2\pi}{\lambda}\right)^3 \,.
\end{equation}
First, we express $\chi_0$ in terms of the detuning and of the
Rabi frequency, utilizing the fact that the light-matter
interaction energy in the Lagrangian (1) of Ref.\ \cite{LKP} is
equal to the optical potential,
\begin{equation}
\frac{\epsilon_0\chi_0}{2} E^2 = \epsilon_0 \chi_0 |E^{(+)}|^2 =
-\hbar\,\frac{|\omega_R|^2}{2\Delta} \,.
\end{equation}
According to Eqs.\ (16) and (36) of Ref.\ \cite{LKP} the light
intensity is
\begin{equation}
I = 2\epsilon_0 c\, |E^{(+)}|^2 \,.
\end{equation}
Consequently, we obtain
\begin{equation}
\gamma_L= \frac{3}{16}\,\Gamma\,
\left|\frac{\omega_R}{2\Delta}\right|^2 \,.
\end{equation}
The upper-state excitation of a far-detuned two-level atom is
$|\omega_R/(2\Delta)|^2$. The incident light excites the atoms,
and the subsequent spontaneous decay gives rise to Rayleigh
scattering, being the principal absorption mechanism. Therefore,
the calculated backaction rate is $3/16$ of the absorption rate.

Even in the limit of extremely far detuning, absorption is still
stronger than the quantum backaction, as can be seen from the
comparison of the two different rates. Consequently, residual
absorption sets indeed the limit of dispersive imaging
\cite{Comment}.

\end{document}